\documentclass[10pt]{iopart}
\begin{document}
\title{The Nishimori line and Bayesian Statistics}

\author{Yukito Iba }

\address{The Institute of Statistical Mathematics,
4-6-7 Minami-Azabu Minato-ku,Tokyo 106-8569,Japan, 
 iba@ism.ac.jp }

\jl{1}

\begin{abstract}
``Nishimori line'' is a line or hypersurface in the
parameter space of systems with quenched disorder,
where simple expressions of the averages of
physical quantities over the quenched random variables are obtained.
It has been playing
an important role in the theoretical studies of the random
frustrated systems since its discovery around 1980.
In this paper, a novel 
interpretation of the Nishimori line 
from the viewpoint of statistical information processing 
is presented. Our main aim is
the reconstruction of the whole theory of the Nishimori
line from the viewpoint of Bayesian statistics, or, almost equivalently,
from the viewpoint of the theory of error-correcting codes. As a byproduct
of our interpretation, counterparts of the Nishimori line in
models without gauge invariance are given. We also discussed
the issues on  the ``finite temperature decoding'' 
of error-correcting codes in connection with our theme and
clarify the role of gauge invariance in this topic.
\end{abstract}

\submitted

\section{Introduction}

There are not many rigorous results that are useful for the study of random frustrated 
systems. Among them, theorems related to the Nishimori line of random spin models 
form an important family. There have been many papers
\cite{Nishi1,Nishi2,Nishi3,Nishi4,Ozeki} about the Nishimori line after the seminal 
paper \cite{Nishi1} of Nishimori.
There is, however, still a mystery about the Nishimori line, i.e., its 
physical meaning and the motivation behind the proof are not yet clear.

The purpose of this paper is to provide a novel interpretation on the Nishimori line 
from the viewpoint of statistical information processing, more specifically from 
Bayesian statistics or from the coding theory.
Our interpretation has two advantages. First it gives an interesting example of an 
unexpected relation between two different areas, rigorous arguments in the 
statistical physics and Bayesian statistics, and 
elucidates the meaning of the trick in the derivation of the Nishimori line.
Secondly it gives some new results on the analog of the Nishimori line without {\it 
gauge invariance} in the sense of Toulouse \cite{Toulouse}.

Our arguments are closely related to the works on the ``the optimality of 
finite-temperature decoding'' of error-correcting codes 
\cite{Rujan,Nishicode,Nishigauge,Sourlas2} . In fact, some of our results are 
essentially given in Sourlas \cite{Sourlas2}. In these works, however, 
finite-temperature decoding is discussed with spin glass theory. On the other hand, our aim 
here is reverse the direction of the arguments and discuss the whole theory of the 
Nishimori line from the viewpoint of statistical science. We will discuss further 
on the finite-temperature decoding in the last section.

In this paper, we make efforts to give a self-contained description of this material. 
No special knowledge on Bayesian statistics, error-correcting codes and gauge 
invariance of spin glass is assumed.

\section{Bayesian Framework}
\label{sec2}

In this section, we give basic notions and terminology of Bayesian statistics. We 
also discuss identities and inequalities that naturally arise from the Bayesian 
framework. 
Although the motivation for these
formulas as well as their proofs are quite simple, they are essential in
the derivation of the properties of the Nishimori line.

Let us assume that our data $y$ is generated by a probability distribution $p(y|x)$, 
which is parameterized by the value of an unknown variable $x$.
In the Bayesian framework, we also assume that the parameter $x$ is, in itself, a 
random sample from a {\it prior distribution} $\pi(x)$.
With these assumptions, the probability distribution of the parameter $x$ conditioned 
on given data $y$ is
\begin{equation}
\label{Bayes}
{ p(x|y)= \frac{p(y|x)\pi(x)}{\sum_x p(y|x)\pi(x) }}_{.}
\end{equation}
Here $\sum_x$ means the summation or integral over the possible values of $x$. This 
distribution, the {\it posterior distribution}, is the source of knowledge with given 
data $y$ in the Bayesian formalism.

Similar formalism is also used in seemingly different branch of the information 
science, the theory of error-correcting codes.
Consider a noisy channel and a set of messages.
We encode and send a message $x$ through the noisy channel and someone at the other 
end of the channel tries to infer the original message $x$ from the output $y$. If 
we assume that the probability $p(y|x)$ of an output $y$ with the input $x$ and the 
distribution $\pi(x)$ of the average frequencies of input messages, the conditional 
probability $p(x|y)$ of an input $x$ with the output $y$ is given by (\ref{Bayes}).
Note that the probability $p(y|x)$ represents the coding scheme as well as the noise 
of the channel in this formalism.

We will introduce notations that indicate the averages over different types of 
distributions.  Here the symbol $A(x)$ denotes a function of the parameter $x$ and 
$B(y)$ denote a function of the data $y$ . First we define the average over the prior 
distribution of $x$,
\begin{equation}
\label{priorA}
[ A(x) ]_{\pi(x)} = \sum_x A(x) \pi(x)_{.}
\end{equation}
We also define the average over the posterior distribution of $x$,
\begin{equation}
\label{postA}
\langle A(x) \rangle_{p(x|y)}  = \sum_x A(x) p(x|y)_{.}
\end{equation}
Finally we define the average over the probability distribution $p(y|x)$ of data 
$y$ with the given parameter $x$,
\begin{equation}
\label{noiseA}
[ B(y) ]_{p(y|x)}  = \sum_y B(y) p(y|x)_{.}
\end{equation}
These notations are not common in the literatures on Bayesian statistics. They are 
introduced to contrast the analogy to the statistical physics of systems with quenched 
disorder. Later we show that the averages $[ \ ]$ correspond to the quenched average 
over the configuration of the impurities and $\langle \ \rangle$ corresponds to the 
thermal average.

Let us consider  relations among these averages. 
First we note that an identity
\begin{equation}
\label{idA}
[ \langle A(x^\prime) \rangle_{p(x^\prime|y)} ]_{p(y|x)} ]_{\pi(x)}
=
{ [ A(x) ]_{\pi(x)}}_{,}
\end{equation}
holds. 
The posterior average $\langle A(x^\prime) \rangle_{p(x^\prime|y)}$ 
can be regarded as an  estimate of  $A(x)$ from the data $y$. It is a random variable 
dependent on $y$ and  the identity (\ref{idA}) shows that
the average of it over the possible
values of the data $y$ and the original parameter $x$ coincides 
with the prior average $[ A(x) ]_{\pi(x)}$. 
The proof of the formula (\ref{idA}) is straightforward.
When we substitute the left hand side of (\ref{idA}) for the definition of the averages 
(\ref{postA}), (\ref{noiseA}), (\ref{priorA}), we obtain the expression,
\begin{equation}
\fl
[ \langle A(x^\prime) \rangle_{p(x^\prime|y)} ]_{p(y|x)} ]_{\pi(x)}
=
\sum_x \sum_y \frac{\sum_{x^\prime} A(x^\prime)p(y|x^\prime)\pi(x^\prime)}
{\sum_{x^\prime} p(y|x^\prime)\pi(x^\prime)} \cdot
p(y|x)\pi(x)_{.}
\end{equation}
By changing the order of the summation and a dummy index, we can show that the factors 
$\sum_{x} p(y|x)\pi(x)$ in the numerator and denominator cancel each other. Using 
$\sum_y p(y|x^\prime)=1$ and  $\sum_{x^\prime} A(x^\prime) \pi(x^\prime) = 
{[ A(x) ]_{\pi(x)}}$, the proof of (\ref{idA}) is completed.

It is easy to generalize (\ref{idA}) to an identity
\begin{equation}
\label{idC}
[ \langle C(x^\prime,y) \rangle_{p(x^\prime|y)} ]_{p(y|x)} ]_{\pi(x)}
=
{ [[ C(x,y) ]_{p(y|x)}]_{\pi(x)}}_{.}
\end{equation}
Here $C(x,y)$ is a function of the data (the output of the channel) $y$ as well as 
the parameter $x$. 
The proof of the relation (\ref{idC}) is essentially the same as that
of (\ref{idA}). The only difference from (\ref{idA}) is that the average 
$[ \ ]_{p(y|x)}$ in the right hand side cannot be removed.

In these arguments, we assume that the ``true'' distributions $p(y|x)$ and 
$\pi(x)$ behind given data are exactly known. They are, however, often unknown in 
a real world example.  A way to fill this gap is to include ``hyper parameters'' 
$\alpha$ and $\gamma$ in the expression of $p(y|x)$ and $\pi(x)$ and estimate them 
from the data. Hereafter we use the notation $p_\alpha(y|x)$ and $\pi_\gamma(x)$ to 
indicate the distributions that contain hyperparameters. An approach to estimate 
hyperparameters $\alpha$ and $\gamma$ from the data $y$ is the minimization of a 
free-energy-like quantity,
\begin{equation}
\label{ABIC}
F(\alpha,\gamma) = - \log \sum_x p_\alpha (y|x) \pi_\gamma (x)_{.}
\end{equation}
Note that the procedure based on the {\it marginal likelihood} $\sum_x p_\alpha (y|x) 
\pi_\gamma (x)$ is successfully used by the many authors in practical problems. It 
is called by a lot of different terms, say, the maximization of {\it type II 
likelihood} \cite{Good,Berger}, the minimization of {\it ABIC} \cite{Akaike}, the 
maximization of {\it evidence} \cite{Mackay1,Mackay2}, and, simply, the maximization 
of the likelihood of $\alpha$ and $\gamma$ \cite{Kitagawa, 
Geman87,Devijver}~\footnote{When the expression of the probability $p(y|x)$ is 
considered
as the function of $x$ with given data $y$, it is called ``likelihood of
the parameter $x$''. This terminology is preferred by the non-Bayesians
who do not treat the parameter $x$ as a random variable, but is also
used by Bayesians. The mixture distribution $\sum_x p_\alpha (x|y)
\pi_\gamma (x) $ at the right hand side of  (\ref{ABIC}) can be regarded
as the likelihood of the hyperparameters $\alpha$ and $\gamma$.}.

At the moment, we assume that the form of the distribution $\pi_\gamma (x)$ and 
$p_\alpha(y|x)$ is correctly known except the values of the hyper parameters.
Even in this case, the hyperparameters $(\alpha, \gamma)$ that maximize (\ref{ABIC}) 
are random variables dependent on the data $y$ and they fluctuate around the true 
values $(\alpha_0,\gamma_0)$ of $(\alpha,\gamma)$.
However,  $(\alpha, \gamma)$ that minimize the average of $F(\alpha, \gamma)$ over 
the true distribution $\pi_{\gamma_0} (x)$ and $p_{\alpha_0}(y|x)$ coincides with 
the true value $(\alpha_0,\gamma_0)$, i.e, the inequality
\begin{equation}
\label{maxA}
[ [ F(\alpha_0,\gamma_0) ]_{p_{\alpha_0} (y|x)} ]_{\pi_{\gamma_0} (x)}
\leq
[ [ F(\alpha,\gamma) ]_{p_{\alpha_0} (y|x)} ]_{\pi_{\gamma_0} (x)}
\end{equation}
holds for any value of $\alpha$ and $\gamma$.

If the right hand side of (\ref{maxA}) is a sufficiently smooth function of 
$(\alpha,\gamma)$, the derivatives of $F$ at 
$(\alpha,\gamma)=(\alpha_0,\gamma_0)$ should be zero. For example,
the following relations are direct consequences of (\ref{maxA}).
\begin{equation}
\label{av1}
 \left. [ [ \, \frac{\partial}{\partial \alpha} F(\alpha,\gamma)
 \, ]_{p_{\alpha_0}(y|x)} ]_{\pi_{\gamma_0} (x)} \right |_{(\alpha,\gamma) 
=(\alpha_0,\gamma_0)} \, =0 {}_.
\end{equation}
\begin{equation}
\label{av2}
\left. [ [ \, \frac{\partial}{\partial \gamma} F(\alpha,\gamma)
 \, ]_{p_{\alpha_0} (y|x)} ]_{\pi_{\gamma_0} (x)}
\right |_{(\alpha,\gamma)=(\alpha_0,\gamma_0)} \, =0_{.}
\end{equation}
Here, the derivatives $\partial F/ \partial\alpha$ and  $\partial F/ \partial\gamma$
should be interpreted as the derivatives $\partial F/ \partial\alpha_k$ and  
$\partial F/ \partial\gamma_k$ with each component of $\alpha=\{\alpha_k\}$ and 
$\gamma=\{\gamma_k\}$, 
when $\alpha$ and $\gamma$ are vectors with more than one
components.  
The conditions on the second derivatives are also derived from (\ref{maxA}) by using 
positive semi-definiteness of the Hessian, say,
\begin{equation}
\label{H1}
\left. [ [ \, \frac{\partial^2}{\partial^2 \alpha} F(\alpha,\gamma)
 \, ]_{p_{\alpha_0} (y|x)} ]_{\pi_{\gamma_0}(x)} \right 
|_{(\alpha,\gamma)=(\alpha_0,\gamma_0)} \, \geq 0_{,}
\end{equation}
which ensure that $(\alpha_0,\gamma_0)$ is 
a relative minimum of (\ref{maxA}).

A simple way to prove (\ref{maxA}) is the use of the Gibbs inequality,
\begin{equation}
\label{GibbsI}
\sum_z P(z) \log \frac{Q(z)}{P(z)} \leq 0
\end{equation}
where $P(z)$ and $Q(z)$ are arbitrary functions that satisfy the relations
$0 \leq P(z),Q(z) \leq 1$ and $\sum_z Q(z)= \sum_z P(z)=1$.
If we set $P(y)=\sum_x p_{\alpha_0} (y|x)\pi_{\gamma_0} (x)$ and $Q(y)=\sum p_\alpha 
(y|x) \pi_\gamma (x) $, it is easy to verify the requirement of the Gibbs inequality 
(\ref{GibbsI}). Then it follows that, for any $\alpha$ and $\gamma$,
\begin{equation}
\left [ \left [ \, \log
\frac{\sum_x p_\alpha (y|x) \pi_\gamma (x)}
       {\sum_x p_{\alpha_0}(y|x) \pi_{\gamma_0} (x)}
\, \right ]_{p_{\alpha_0} (y|x)} \right ]_{\pi_{\gamma_0} (x)}
\leq 0_{.}
\end{equation}
This proves (\ref{maxA}) and its corollaries (\ref{av1}),(\ref{av2}),(\ref{H1}).
We can also prove (\ref{av1}),(\ref{av2}),(\ref{H1}) through direct calculations
similar to that for (\ref{idA}). 

The history of Bayesian statistics is long and complicated.
It originates from the Laplace's works on the ``inverse probability'' and has been 
an archetype of mathematical theories of uncertain objects. Despite different 
interpretations and objections to the use of prior distributions, it is an important 
language in a wide area of the sciences of information processing, say, time-series 
analysis \cite{Akaike, Kitagawa,West}, image restoration 
\cite{Besag1,Besag2,Geman,Geman87,Derin,Marroquin2,Devijver,Winkler}, inference 
with neural networks \cite{Mackay2,Neal}, and artificial intelligence.
An earlier remark on the analogy between Bayesian statistics and statistical 
mechanics is found, for example, in Iba~\cite{Iba1}. Sourlas~\cite{Sourlas1} seems 
the first reference that discussed the relation between the coding theory and spin 
glasses. We also refer recent works of
Bruce and Saad~\cite{Bruce}, Pryce and Bruce~\cite{Pryce}, Tanaka and 
Morita~\cite{Tanaka}, Kabashima and Saad~\cite{KS97,KS98}, 
{Dress et al.}~\cite{Dress}, Opper and Winther~\cite{Opper}, which treat the 
relation between 
statistical mechanics and Bayesian statistics (or error-correcting codes).

\section{The Nishimori Line}
\label{secN}

Now we discuss the relation between the results in the previous section and the 
Nishimori line of spin glasses. To see this in the simplest case of $\pm J$ Ising 
spin glass, we set the distributions as follows:
\begin{equation}
\label{pair0}
p_\alpha (y|x) = \frac{1}{Z_{\alpha}} \exp(-E_\alpha (x,y))_,
\end{equation}
\begin{equation}
- E_\alpha (x,y) = \alpha \sum_{(i,j)} y_{ij} x_i x_j {}_,
\end{equation}
\begin{equation}
\label{pair1}
Z_\alpha = \sum_y \exp(-E_\alpha (x,y)) =
(\exp(\alpha)+\exp(-\alpha))^M {}_,
\end{equation}
and
\begin{equation}
\label{pair2}
\pi(x)= \frac{1}{2^N} \, \mbox{(the uniform distribution)}_{,}
\end{equation}
where each of the component $x_i \, ( i \in \{1..N\})$ of the parameter $x$ takes 
the value of $\pm 1$.  The component $x_i$ is defined on the vertices $i$ of a graph 
$G$, say a square lattice or a random network, of degree 
$N$. The data $y=\{y_{ij}\}$ is defined on the edges $(i,j)$ of $G$ and the summation 
$\sum_{(i,j)}$ runs over them. We denote the number of the edges
of $G$ as $M$, which is also the number of the data.

This probability $p_\alpha(y|x)$ corresponds to a binary symmetric
channel where a set $\{y^{in}_{ij}\} \, ((i,j) \in G)$ of the pair product 
$y^{in}_{ij}=x_i x_j$ of the inputs is sent as an error-correcting code 
\cite{Sourlas1,Sourlas2,Nishicode,KS97}. Here, ``binary symmetric'' means that the 
output of the channel $y_{ij}$ is given by the formula
\begin{eqnarray}
y_{ij}   = + y_{ij}^{in}   &  \mbox{ with probability} \  & 1-q
 \nonumber \\
y_{ij}   = - y_{ij}^{in}   &  \mbox{ with probability}   \ &  q {}_.
\end{eqnarray}
If we assume that the data $y_{ij}$ is generated by $p_{\alpha_0}(y|x)$ and $\pi(x)$
defined by (\ref{pair0}) and (\ref{pair2}), the noise $q$ of the channel 
is related to the hyperparameter $\alpha_0$ by
\begin{equation}
q=\frac{\exp(-\alpha_0)}{\exp(\alpha_0)+\exp(-\alpha_0)} \, {}_.
\end{equation}
Although ``pair product code'' $y^{in}_{ij}=x_i x_j$ defined in the above looks 
rather artificial one, recent works \cite{KS97, KS98} on error-correcting codes 
suggest its generalization might have practical importance~\footnote{
Another interesting interpretation of the probability $p_\alpha(y|x)$  is given by 
a problem \cite{Iba2} that arises in the analysis of social network data \cite{Wang}.
With this interpretation, the index $i$ indicate a person and the binary variable
$y_{ij} \in \{ \pm 1 \} \, (y_{ij}=y_{ji})$ indicates a social relation between 
persons $i$ and $j$, say, whether they have an acquaintance or not. Each person is 
assumed to subject to one of the social groups A and B, and the problem is to infer 
the group structure from the data $\{y_{ij}\}$.  We set the indicator $x_i=1$ when 
$i \in A$ and $x_i=-1$ when  $i \in B$ and assume the following property:
\begin{quote}
If a pair of the persons $i$ and $j$ is in the same social group, $y_{ij}=1$ with 
a probability $q$ and $-1$ with a probability $1-q$. Else if they are in different 
groups, $y_{ij}=1$ with probability $q^\prime$ and $-1$ with a probability $1-
q^\prime$.
\end{quote}
Then we get the probability $p_\alpha(y|x)$  in the text as a special case where 
$q^\prime=1-q$.}.

The posterior distribution of the model with data $\{y_{ij}\}$ and 
hyperparameter $\alpha$ is
\begin{equation}
\label{posSG}
p_\alpha (x|y) = \frac{1}{Z_{pos}} \exp(-E_\alpha (x,y)) {}_,
\end{equation}
\begin{equation}
Z_{pos} = \sum_x \exp(-E_\alpha (x,y)) {}_.
\end{equation}
This is the Gibbs distribution of a random bond Ising model
with coupling constants $\{y_{ij}\}$ defined on the graph $G$.
The derivatives of the function $F$ defined by (\ref{ABIC}) are
\begin{equation}
\label{eSG}
\frac{\partial}{\partial \alpha} F(\alpha,\gamma)
= - \frac{1}{\alpha} \langle  E_\alpha (x^\prime,y) \rangle_{p_\alpha(x^\prime|y)} 
- M \cdot \tanh \alpha
\end{equation}
where $\langle \ \rangle_{p_\alpha(x|y)}$ indicates the canonical average with the 
energy $E(x,y)$ (Here and hereafter, we assume that we are working at the unit 
temperature $T=1$ and $\alpha$ is treated as a (hyper)parameter of the model but not 
the temperature.). The term $M \cdot \tanh \alpha$ comes from the derivative of the 
logarithm of the normalization factor \mbox{$(\exp(h)+\exp(-h))^M$} of the 
probability $p_\alpha(y|x)$.

In general, a misspeficification of hyperparameter $\alpha$ in (\ref{posSG}) and 
(\ref{eSG})
is possible.
If we assume that we know the ``true'' value $\alpha_0$
used in the generation of the data $\{y_{ij}\}$
and set  $\alpha=\alpha_0$, or, equivalently,
\begin{equation}
\label{Nline}
\frac{\exp(-\alpha)}{\exp(\alpha)+\exp(-\alpha)} = q \, {}_,
\end{equation}
in the formulas, we have an identity on the energy
\begin{equation}
\label{sg1}
- [ \langle  E_\alpha  (x^\prime,y) \rangle_{p_\alpha (x^\prime | y)} ]_{p_\alpha 
(y|x)} ]_{\pi(x)} = \alpha M \cdot \tanh \alpha  {}_.
\end{equation}
from the identity (\ref{av1}). 

So far, the average over the bond randomness
\begin{equation}
[ \, \, [ \, \cdots \, ]_{p_\alpha (y|x)} \, ]_{\pi(x)}
\end{equation}
has a rather complicated structure. There are two steps in the generation of the 
quenched random variables $\{y_{ij}\}$, which are described by
$\pi(x)$ and $p_\alpha (y|x)$ respectively. In this particular case, we can simplify 
it using {\it gauge invariance} of the problem. The gauge transformation group of
this model is defined by the family of
transformations,  
\begin{equation}
\label{GX}
V_z: \{y_{ij}\} \rightarrow \{ z_i \cdot y_{ij} \cdot 
z_j \} 
\end{equation}
\begin{equation}
\label{GY}
U_z: \{x_i \} \rightarrow \{ x_i \cdot z_i \}
\end{equation}
parameterized by $z=\{z_i\}, z_i \in \{\pm 1 \}$.
This set of transformations consist of 
one-to-one onto-mappings (permutations) of their domain and satisfy a transitive 
property, i.e.,
there exists $z$ with which $U_z(x)=x^\prime$ 
for any pair of $x$ and $x^\prime$ in the domain.
It is easy to show the following relations:
\begin{equation}
\label{lgauge}
p_\alpha(U_z(x)|V_z(y))= p_\alpha(x|y)
\end{equation}
\begin{equation}
\label{pigauge}
\pi(U_z(x))= \pi(x)
\end{equation}
\begin{equation}
\label{posgauge}
p_\alpha(V_z(y)|U_z(x))= p_\alpha(y|x) {}_,
\end{equation}
\begin{equation} 
E_\alpha (U_z(x),V_z(y))= E_\alpha(x,y) {}_.
\end{equation}
They are an
example of the gauge invariance (or gauge covariance) in a random spin system 
(\cite{Toulouse,Fradkin,Nishi1,Nishi2}, see \cite{Ozeki} for a comprehensive 
treatment
with applications to the Nishimori line.).
By these formulas, we can show that
the left hand side of the expression (\ref{sg1}) is written as
a simpler average
\begin{eqnarray}
\label{ggg}
\fl \lefteqn{[ [ \langle  E_\alpha  (x^\prime,y) \rangle_{p_\alpha (x^\prime | 
y)} ]_{p_\alpha (y|x)} ]_{\pi(x)}} \nonumber \\
\fl & & = \sum_x \sum_y \sum_{x^\prime}
E_\alpha  (x^\prime,y) \cdot
p_\alpha (x^\prime | y) \cdot
p_\alpha (y|x) \cdot \pi(x)
\nonumber  \\
\fl  & & = \sum_x \sum_y \sum_{x^\prime}
E_\alpha (U_z(x^\prime),V_z(y)) \cdot
p_\alpha (U_z(x^\prime) | V_z(y)) \cdot
p_\alpha (V_z(y)|U_z(x)) \cdot \pi(x) \nonumber \\
\fl & &
= \sum_y \sum_{x^\prime}
E_\alpha (x^\prime ,y) \cdot
p_\alpha (x^\prime |y) \cdot
p_\alpha (y|x^*)  \nonumber \\
\fl & & = [ \langle  E_\alpha  (x^\prime,y) \rangle_{p_\alpha (x^\prime | 
y)} ]_{p_\alpha (y|x^*)} 
\end{eqnarray}
where $x^*$ is a ferromagnetic state 
$\{x^*_i \} \, ( \forall i \, x^*_i=1) $ and
$z$ is a function of $x$ that satisfy $x^*=U_z(x)$. 
The existence
of such $z$ is secured by the transitive property and the
change of the dummy index, say, from $V_z(y)$ to $y$
in the summation $\sum_y$, is justified 
by the one-to-one onto property of the gauge transformations (\ref{GX}),(\ref{GY})
respectively.

Here and hereafter, we denote the average over the distribution
\begin{equation}
\label{free}
p_\alpha(y|x^*) = \frac{\exp( \alpha \sum_{(i,j)} y_{ij}) }{Z_\alpha}
\end{equation}
as $[ \ \ ]_q$, where the relation between $q$ and $\alpha$ is defined in (\ref{Nline}). 
It is easy to see a component $y_{ij}$ of sample from the distribution (\ref{free}) 
is mutually independent samples from the distribution
\begin{equation}
\label{random}
\Pr(y_{ij}) = q \cdot \delta(y_{ij}-1) + (1-q) \cdot \delta(y_{ij}+1)_{.}
\end{equation}
By using (\ref{ggg}) and these notations, the formula (\ref{sg1}) is reduced to
the identity
\begin{equation}
\label{idE}
- [ \langle E(x,y) \rangle_{p_\alpha(x|y)} ]_q = \alpha M \cdot \tanh \alpha
\end{equation}
on the average of the energy of $\pm J$ spin glass with the coupling
constants $\{y_{ij}\}$ from the distribution (\ref{random}).
{It is nothing but a result reported in the first paper
\cite{Nishi1} on the Nishimori line.}

The relation  (\ref{Nline}) 
between (hyper)parameter $\alpha$ in the canonical average and the noise level $q$ in 
the quenched average 
is essential and known as the definition of the {\it Nishimori line} of the model. 
In our derivation, it arises from the condition $\alpha=\alpha_0$
in the formulas (\ref{maxA}) and (\ref{av1}).
This means that the model $p_\alpha(y|x)$ assumed in the analysis of the data
coincides with  the ``true''  probability $p_{\alpha_0}(y|x)$ used in the 
generation of the data. 
In terms of the coding theory, it corresponds to the situation where the decoder knows 
exactly the property of the channel, the coding, and the relative frequencies of the 
possible messages.
It is rather surprising that the notion of the Nishimori line~\footnote{
In general cases, where more than one (components of) hyperparameters are contained 
in the model, it  is actually ``Nishimori hypersurface''. The term {\it Nishimori 
temperature} is also used by statistical physicists. It seems, however, not adequate 
terminology in the context of information processing, because the notion of 
temperature has no specific meaning in the problems in the statistics and the coding 
theory.}, which is introduced without any background of statistical information 
processing, has such a clear interpretation from our point of view.

A similar argument with the substitution of the second derivative 
$\frac{ \partial^2 }{ \partial^2 \alpha } F(\alpha,\gamma)$ of  $F$ into the 
expression (\ref{H1}) leads to an inequality
\begin{equation}
\label{sg2}
[ [ \, \langle  E_\alpha^2 (x^\prime,y)   \rangle_{p_\alpha (x^\prime | y)} -
\langle  E_\alpha(x^\prime,y)    \rangle_{p_\alpha (x^\prime | y)}^2
\, ]_{p_\alpha (y|x)}]_{\pi(x)} \, \leq  \frac{\alpha^2 M}{\cosh^2 \alpha}  {}_.
\end{equation}
With gauge invariance of the model, we can derive an inequality
\begin{equation}
\label{idE2}
[ \langle E_\alpha (x,y)^2 \rangle_{p_\alpha(x|y)} - \langle E_\alpha (x,y) 
\rangle^2_{p_\alpha(x|y)}]_q \leq \frac{\alpha^2 M}{\cosh^2 \alpha}
\end{equation}
from (\ref{sg2}). This is an inequality on the fluctuation of the energy (the specific 
heat) on the Nishimori line, which is also discussed in \cite{Nishi1}.
Some of the other relations that hold on the Nishimori line of the model
is derived from the identity (\ref{idC}) and the gauge invariance of the model. 
For example, the distribution of the internal fields at the vertex $i_0$ \cite{Nishi3} 
is reproduced, when we set $C(x,y)=\sum_j y_{i_0 j} x_j$, where $j$ 
runs over the set of vertices neighboring to $i_0$, i.e., $(i_0,j) \in G$.
The expression of the gauge invariant correlation function
\cite{Nishi1} is also derived, when we set $C(x,y)= x_k \cdot ( \Pi_{(i,j) \in \Gamma } 
\, \, y_{ij} ) \cdot x_l$, where $\Gamma$ denotes a path that connects the vertices 
$k$ and $l$.

Here we discuss a statistical model defined by (\ref{pair0}) and
(\ref{pair2}), which leads to the Nishimori line of the $\pm J$ spin glass
model. Our argument is, however, general and can be applied to the Nishimori line 
of other models, say, spin glasses with a Gaussian distribution
of the coupling \cite{Nishi2}, models with multiple spin interactions 
\cite{Sourlas1,Nishicode,Ozeki,Dress,KS97,KS98}, 
and the gauge glasses \cite{Ozeki,Nishigauge}. For each model, we can consider the
corresponding statistical model (or a noisy channel) and derive the
properties of the Nishimori line from the relations (\ref{av1}), (\ref{H1}),
(\ref{idC}) with additional arguments on the gauge invariance.
For example, we consider the following problem of statistical inference:
\begin{quote}
There are a collection of the phase variable $\{x_i\} \, (x_i \in [0,2\pi) \, )$ on 
the vertices of a graph $G$, say, a two-dimensional lattice. 
The numbers of the vertices and edges in $G$ are $N$ and $M$, respectively.
We assume that
the difference $y^{in}_{ij}  = x_i - x_j $  of the parameters $x_i $ and $x_j$ is 
observed for each edge $(i, j)$ in $G$ with a probabilistic error $\eta_{ij}$.  The 
problem is to infer the original values of $\{x_i\}$ from the data $\{y_{ij}\}$ with 
the assumption that
$y_{ij}=y_{ij}^{in}+\eta_{ij}$.
\end{quote}
Such a problem might have practical importance in the analysis
of the data from optical measurements where differences of the phase
between neighboring points are observed with noise.
If we assume that the magnitudes of the noise $\{\eta_{ij}\}$ are mutually independent 
and obey a von Mises distribution, a correspondence
of a Gaussian distribution on a circle, the probability is
\begin{equation}
p_\alpha (y|x) = \frac{1}{Z_{\alpha}} \exp(-E_\alpha (x, y))_, \nonumber
\end{equation}
\begin{equation}
- E_\alpha (x, y) = \alpha \sum_{(i,j)} \cos(x_i-x_j-y_{ij}) {}_, \nonumber
\end{equation}
\begin{equation}
\label{gauge1}
Z_\alpha = \int \Pi_i \, dx_i \, \, \, p_\alpha(y|x) 
= (2\pi I_0(\alpha))^M {}_,
\end{equation}
where $I_0$ indicates the modified Bessel function.
We also assume the uniform prior
\begin{equation}
\label{gauge2}
\pi(x) dx= (\frac{1}{2\pi})^N \cdot \Pi_i \, dx_i , \qquad x_i \in [0,2\pi) {}_.
\end{equation}
From this setting,
we can derive the results on the Nishimori line of the {\it gauge glass} 
\cite{Ozeki,Nishigauge}
with a method that is similar to that for the $\pm J$ spin glass.

\section{What is New ?}

Now, careful readers may ask {\it what is really new} in our approach.
Once the expression 
\begin{equation}
\label{sg1g}
\fl
\sum_x
\sum_y
\left (
\frac {
\sum_{x^\prime}
( \sum_{(i,j)} y_{i j} x_i^\prime x_j ^\prime) \cdot
\exp( \alpha \sum_{(i,j)} y_{ij} x_i^\prime x_j^\prime )
}
{
\sum_{x^\prime}
\exp( \alpha \sum_{(i,j)} y_{ij} x_i^\prime x_j^\prime )
}
\cdot
\frac{\exp( \alpha \sum_{(i,j)} y_{ij} x_i x_j) }{Z_\alpha} 
\right )
\end{equation}
of the left hand side of (\ref{idE}) is derived by
the gauge invariance from that of (\ref{sg1}), it is not difficult to
show the relation (\ref{idE}) by direct inspection 
of the expression.
If we combined these steps of the proof in this order, it is nothing but
a conventional proof  \cite{Nishi2} of the property of the Nishimori line. The same 
is
true for the derivation with the identity (\ref{idC}). 
In this sense, our argument is not a re-derivation of the Nishimori line
but a {\it reformulation} or {\it re-interpretation} of the original
derivation.

There are, however, two major advantages of our approach.
First, our interpretation elucidates the meaning of the Nishimori line.
It is a line on which we make inference (or decoding) using the ``true'' probability 
structure that generates the data (or codes).  This coincidence of the encoding and 
decoding scheme gives drastic simplifications of the averages of various kind of 
physical quantities. This interpretation also explains why there are two different 
variables $x$ and $x^\prime$ with the same type in the expression (\ref{sg1g}). The 
variable $x$ corresponds to the original value of the parameter (the input message) 
and $x^\prime$ represents an inference on it.  The expression (\ref{sg1g}), in itself, 
has clear meaning as the relation (\ref{sg1}) between averages.
On the other hand, in the conventional derivation \cite{Nishi2} of the Nishimori line, 
the insertion of the variable $x$ to the left hand side of (\ref{idE})  looks a rather 
artificial procedure 
and the expression (\ref{sg1g}), which  is defined in the configuration space 
enlarged  by a gauge transformation, seems to have  no definite meaning. This lacks 
of the interpretation is a reason why the derivation of the Nishimori line looks 
somewhat mysterious, even though the manipulation of the formula required in the proof 
is quite simple and elegant. We believe that our interpretation will contribute to 
make this point clear. Another interesting point of our argument is that it gives 
a novel interpretation to the identity (\ref{idE}) of the energy and the inequality 
(\ref{idE2}). 
They are essentially necessary conditions that the average of the marginal likelihood 
takes the maximum value on the Nishimori line.

The second advantage of our approach is that it suggests the existence of the 
correspondence of the Nishimori line in the models {\it without gauge invariance.}  
The relations (\ref{av1}),(\ref{H1}), and (\ref{idC}), which are used in the 
derivation of the Nishimori line, are obtained without the gauge invariance of the 
models. Thus, we can prove the identity of the energy, inequality of the specific 
heat, and the expression of the distribution of internal fields etc., which hold on 
the ``Nishimori line'' of the models without gauge invariance.
For example, we consider a Bayesian model
\begin{equation}
\label{BSCN}
p_\alpha (y|x) = \frac{1}{Z_{\alpha}} \exp(-E_\alpha (x,y))_,
\end{equation}
\begin{equation}
- E_\alpha (x,y) = \alpha \sum_i y_i x_i {}_,
\end{equation}
\begin{equation}
Z_\alpha = \sum_y \exp(-E_\alpha (x,y)) =
(\exp(\alpha)+\exp(-\alpha))^N {}_,
\end{equation}
and
\begin{equation}
\label{IGG}
\pi_\gamma(x)= \frac{1}{Z_\pi} \exp(-E_\gamma (x,y))_,
\end{equation}
\begin{equation}
- E_\gamma (x) = \gamma \sum_{(i,j)} x_i x_j {}_,
\end{equation}
\begin{equation}
Z_\gamma = \sum_x \exp(-E_\gamma (x)) {}_.
\end{equation}
In this case, we assume that the unknown parameters $\{x_i\}$ 
and the data $\{y_i\}$ are defined on the vertices of 
a graph $G$ with the degree $N$. When $G$ is a two or three dimensional lattice, this 
model corresponds to an image restoration problem
with a prior knowledge on images that is well described by
the Ising prior (\ref{IGG})(For image restoration 
with Ising and Potts priors, see the references 
\cite{Besag1,Besag2,Geman,Marroquin2,Winkler,Pryce,Tanaka}.).
The posterior distribution of the model is
\begin{equation}
\label{IGGPOS}
p_{\alpha\gamma} (x|y) = \frac{1}{Z_{\alpha\gamma}} 
\exp(-E_{\alpha\gamma} (x,y)) {}_,
\end{equation}
\begin{equation}
\label{RFIME}
- E_{\alpha\gamma} (x,y) = \alpha \sum_i y_i x_i + \gamma \sum_{(i,j)} x_i x_j {}_,
\end{equation}
\begin{equation}
Z_{\alpha\gamma} = \sum_x \exp(-E_{\alpha\gamma}(x,y)) {}_.
\end{equation}
This is the Gibbs distribution of an Ising model with an inhomogeneous
external field $\{ h \cdot y_i \}$.
 
Let us consider the cases where the data generation mechanism is exactly 
described by the probabilities (\ref{BSCN}) with $\alpha=\alpha_0$ 
and (\ref{IGG}) with $\gamma=\gamma_0$, i.e., the pattern of the 
random field $\{y_i\}$ is given by the following process : 
(i)~Generate a sample pattern $\{y_i^{in}\}$ from 
the Gibbs distribution
of the Ising model (\ref{IGG}) with the 
coupling constant $\gamma_0$. (ii)~Flip each component
$\{y_i^{in}\}$ with the probability 
\begin{equation}
q=\frac{\exp(-\alpha_0)}{\exp(\alpha_0)+\exp(-\alpha_0)}
\end{equation}
 (a binary symmetric channel). Then,
we can interpret the posterior (\ref{IGGPOS}) as a Gibbs distribution of
a Random Field Ising Model (RFIM). Note that external fields
on the sites of this model are not 
mutually independent random variables, but correlated with 
a way specified by (i) and (ii). 
The ``Nishimori line'' of this model is defined as a surface where
the parameters $(\alpha_0,\gamma_0)$ in the definition of the quenched randomness
coincide with $(\alpha,\gamma)$ in the canonical average. Equivalently,
\begin{equation}
\frac{\exp(-\alpha)}
{\exp(\alpha)+\exp(-\alpha)} = q{}_,
\end{equation}
\begin{equation}
\gamma = \gamma_0  {}_,
\end{equation}
where $\alpha$ and $\gamma$ are the parameters
in (\ref{RFIME}), and $q$ and $\gamma_0$ are
the parameters in (i) and (ii).
From (\ref{av1})(\ref{av2}) and (\ref{idA}) with $A(x)=x_ix_j$,  we can prove 
identities that holds on the Nishimori line of this model,
\begin{equation}
[[ \, \langle \sum_i y_i x_i \rangle_{p_{\alpha\gamma}(x|y)} 
\, ]_{p_\alpha(y|x)} ]_{\pi_\gamma(x)}
=
N \cdot \tanh \alpha {}_,
\end{equation}
\begin{equation}
[[ \, \langle \sum_{(i,j)} x_i x_j \rangle_{p_{\alpha\gamma}(x|y)} 
\, ]_{p_\alpha(y|x)} ]_{\pi_\gamma(x)}
=
\langle \sum_{(i,j)} x_ix_j \rangle _\gamma^{pure} {}_,
\end{equation}
\begin{equation}
\label{RFIMcr}
[[ \, \langle x_i x_j \rangle_{p_{\alpha\gamma}(x|y)} 
\, ]_{p_\alpha(y|x)} ]_{\pi_\gamma(x)}
=
\langle x_ix_j \rangle _\gamma^{pure} {}_.
\end{equation}
Here and hereafter, the average $\langle \, \cdots \, \rangle _\gamma^{pure}$
is the canonical average with the ``pure'' Ising model with homogeneous 
couplings of the strength $\gamma$ on the same graph $G$.

The expression (\ref{RFIMcr}) shows that the quenched average of the correlation of 
spins  in the RFIM with a correlated random field is just the same as 
that of the corresponding pure Ising model. 
Furthermore, an identity between the order parameters is obtained if we consider a 
set of systems of a fixed boundary condition with which $x_i=1$ for the all
spins at the boundary \cite{Nishi2}. That is, the random field $\{y_i\}$ is assumed 
to be 
generated by the process (i) and (ii) in which the values of the boundary spins are 
kept to 1, and 
the thermal averages $\langle x_i x_j \rangle_{p_{\alpha\gamma}(x|y)}$ and
$\langle x_ix_j \rangle _\gamma^{pure}$ are also taken with this boundary condition.
Assuming that the site $j$ belongs to the 
boundary and the site $i$ is located far from the boundary, the relation
\begin{equation}
[[ \, \langle x_i  \rangle_{p_{\alpha\gamma}(x|y)} 
\, ]_{p_\alpha(y|x)} ]_{\pi_\gamma(x)}
=
m_\gamma^{pure}
\end{equation}
is derived from (\ref{RFIMcr}), where $m_\gamma^{pure}$ is the bulk 
magnetization per spin of the corresponding pure system.

Although these results are dependent on the special features of the
model, similar arguments are applicable in other models
without gauge invariance and leads to
identities and inequalities on the Nishimori line
of the model. An example is provided by the posterior
distribution corresponding to a binary asymmetric channel, which is
already discussed in Sourlas \cite{Sourlas2} in the context of optimal 
decoding. It is related to models with a special type of
site/bond randomness.

There are, however, some intrinsic limitations on the utility of the notion of the 
Nishimori line without gauge invariance. First, we cannot simplify the definition 
of the quenched average
$[ \, \, [ \, \cdots \, ]_{p_\alpha (y|x)} \, ]_{\pi_\gamma (x)}$ on the Nishimori 
line without gauge invariance. Then, the results usually contain a complicated 
quenched average, which often lacks a clear correspondence to that in physical systems. 
The two-stage process (i) and (ii) of the generation of quenched randomness in the 
RFIM just described above is a typical example of this.
Another important remark is that not all of the arguments on the Nishimori line 
with gauge invariance is applicable to the models without gauge invariance.
For example, the identity
\begin{equation}
\label{Order}
[\, \langle x_i^{\prime} \rangle_{p_\alpha(x^{\prime}_i|y)} \cdot \langle 
x_i^\prime  \rangle_{p_{\alpha^\prime} (x^\prime |y)} \, ]_q
=
[\, \langle x_i^\prime  \rangle_{p_{\alpha^\prime}(x^\prime |y)} \, ]_q
\end{equation}
valid for the $\pm J$ spin glass model \cite{Nishi2} has no correspondence in a model 
without gauge invariance. Here $\alpha$
and $q$ is related by the condition (\ref{Nline}) of the Nishimori line and
$\alpha^\prime$ takes an arbitrary value.
The identity (\ref{Order}) is important because the upper bound
\begin{equation}
\label{OrderI}
| [ \, \langle x^\prime _i \rangle_{p_{\alpha^\prime} (x^\prime |y)} \, ]_q |
\leq [ | \langle x_i \rangle |_{p_\alpha (x|y)} ]_q
\end{equation}
of the order parameter is derived from it. If we substitute $C(x,y)= x_i \cdot \langle 
x_i^\prime \rangle_{p_{\alpha^\prime} (x_i^\prime |y)}$ in (\ref{idC}), we can prove 
the relation
\begin{equation}
\label{notOrder}
\fl
[[ \, \langle x_i^{\prime} \rangle_{p_\alpha(x^{\prime}_i|y)} \cdot \langle 
x_i^\prime  \rangle_{p_{\alpha^\prime} (x^\prime |y)} 
\, ]_{p_\alpha(y|x)}]_{\pi_\gamma(x)}
=
[[ \, x_i \cdot \langle x_i^\prime  \rangle_{p_{\alpha^\prime} (x^\prime |y)} 
\, ]_{p_\alpha(y|x)}]_{\pi_\gamma(x)} {}_,
\end{equation}
which apparently corresponds to (\ref{Order}). However, further simplification of 
the right hand side is not possible without gauge invariance. Unfortunately, the 
expression (\ref{notOrder}) gives little information on the shape of the boundaries 
in phase diagram and seems not so useful as (\ref{Order}).

\section{Finite-temperature Decoding}

The notion of the optimality of ``finite temperature decoding'' is introduced to the 
community of statistical physicists by Ruj\'{a}n \cite{Rujan} and discussed by 
Nishimori \cite{Nishicode,Nishigauge} and Sourlas \cite{Sourlas2}. Recently, it 
again draws an attention of the researchers of this area, because the development 
on the statistical mechanics of error-correcting codes \cite{KS97}
enables the quantitative argument of the problem with analytical methods.

Roughly speaking, ``the optimality of finite temperature decoding'' means that the 
estimator that maximize the posterior probability (Maximum A Posteriori estimator) 
is not always the best estimator. The best estimator is dependent on the purpose of 
inference (or decoding) and often defined with averages over the posterior 
distribution. If we call MAP estimator, which is defined as a ``ground state'' of 
the corresponding physical system, the ``zero-temperature decoder'', it is natural 
to call an estimator defined with the posterior averages a ``finite temperature 
decoder'' or ``$T=1$ decoder'' \cite{Sourlas2}.

This fact, however, had been well known in the study of the statistics and pattern 
recognition. For example, Marroquin \cite{Marroquin1}(see also 
\cite{Marroquin2,Winkler}) 
discussed an estimator (``MPM estimator'') in image restoration problems, which is 
just the same as the one proposed by Ruj\'{a}n \cite{Rujan}.  
Moreover, 
it is not the first work that uses the estimator in this field~\footnote{
See, for example, the references \cite{Derin} and Sec 2.4 of \cite{Besag2}.
A recent work on the optimal estimator in image restoration 
is found in Rue \cite{Rue}.}.
General arguments on the optimality of the estimator in the Bayesian framework is 
already found in the textbooks \cite{Text1,TextJ,Text4, Berger} of 
statistics. The branch of statistics that discuss optimal decisions with 
uncertain information is known as {\it statistical decision theory}. 

Here, we will briefly discuss the basic results on optimal estimators. Our treatment 
is not very different from the arguments in Sourlas \cite{Sourlas2} and those in the 
textbooks of statistics. 
It is, however, useful to give a coherent derivation 
with the notations in the earlier sections, because no comprehensive
treatment on this subject seems available in the literature of physics.

To give a formal definition of optimal estimators, we introduce the notion of a {\it 
loss function} $L(x,\hat{x})$ that gives a measure of distance~\footnote{It is {\it not} 
necessary to satisfy the axiom of the distance.}
between the original parameter $x$ and 
an estimate $\hat{x}$ of $x$. 
Then we define 
an optimal estimator $\hat{x}(y)$ for a loss function $L$ as a function 
of $y$ that minimize the expected loss
\begin{equation}
[ [ L(x,\hat{x}(y))]_{p(y|x)}]_{\pi(x)} {}_.
\end{equation}
Here and hereafter we assume that we know exactly about the data generation process
and omit the subscripts that indicate hyper parameters $\alpha, \gamma$ 
in the expressions, say, $p(y|x)$, $\pi(x)$ and $\langle \ \rangle$ (i.e., we set 
the values of the
hyperparameters to their ``true'' values.).
Note that the optimality of an estimator defined here is a very strong notion.
It means that $\hat{x}(y)$ has better or equal average performance against any 
function of the data $y$, provided that the data generation scheme (or the set of 
the channel and the frequencies of the messages) is exactly described by the given 
probability $p(y|x)$ and $\pi(x)$. It is not restricted to the optimality in a series 
of the estimators defined with different hyperparameters or temperatures.

A basic result on optimal estimators is as follows:
\begin{quote}
{\bf Lemma} \\
An optimal estimator $\hat{x}(y)$ for a loss function $L$ is an estimator that 
minimize the posterior average $\langle L(x,\hat{x}(y)) \rangle_{p(x|y)}$ for each 
$y$. 
\end{quote}
This result is a rather obvious one with the principle that the posterior distribution
is the source of the all information from the data.
If we use the identity (\ref{idC}), the formal derivation of the lemma is easy. When 
we set $C(x,y)=L(x,\hat{x}(y))$, the expression (\ref{idC}) gives
\begin{equation}
\label{idL}
[ [ \langle L(x^\prime ,\hat{x}(y)) \rangle_{p(x^\prime|y)}]_{p(y|x)}]_{\pi(x)}
 = [ [ L(x,\hat{x}(y))]_{p(y|x)}]_{\pi(x)} {}_.
\end{equation}
Here we note three observations on the formulas ($\ref{idL}$): \
(a) The estimator $\hat{x}(y)$ is an arbitrary function of $y$ and we can freely 
attribute its value at each $y$. 
(b) The average
 $[ \, \, [ \, \cdots \,]_{p(y|x)} \, ]_{\pi(x)}$ in the left hand side of (\ref{idL}) 
is an average over $y$ with non-negative weights. 
(c) The function $L(x^\prime, \hat{x}(y))$ does not explicitly contains
the original parameter $x$.
By using (a),(b) and (c), we can see that the minimizer of the left side hand of 
$(\ref{idL})$ is the minimizer of the posterior average $\langle L(x,\hat{x}(y)) 
\rangle_{p(x|y)}$ for each $y$. Thus, 
the lemma is proved.

For example, consider the case where the distance $L$ between the binary sequence 
$x=\{x_i\}$ and $\hat{x}=\{ \hat{x}_i \} \, ( x_i, \hat{x}_i \in \{\pm 1\} )$ is 
measured by the overlap $\sum_i \hat{x}_i x_i$ of the pattern, i.e., $L(x,\hat{x})= 
- \sum_i \hat{x}_i x_i$. With this loss function,
\begin{equation}
\label{exopt}
\langle L(x,\hat{x}) \rangle_{p(x|y)} =
 - \sum_i \hat{x}_i(y)  \langle x_i \rangle_{p(x|y)} {}_.
\end{equation}
Then, the optimal estimator $\hat{x}_i (y) \in \{ \pm 1 \}$, which minimize the right 
hand side of (\ref{exopt}), is given by
\begin{equation}
\hat{x}_i (y)=\frac{\langle x_i \rangle_{p(x|y)}}{| \langle x_i \rangle_{p(x|y)} |} 
\, {}_.
\end{equation}
This expression coincides with the result in 
\cite{Rujan,Nishicode,Sourlas2,Marroquin1, Marroquin2,Winkler}. 
Examples of loss functions and the corresponding optimal estimators are shown in the 
Table~1. By using the lemma, we can easily derive them. 

\fulltable{Loss Functions and Corresponding Optimal Estimators. \\
If there are no special comments in the table, a component of parameters $x_i$ \\ and 
its estimate $\hat{x}_i$ are assumed to take their values 
in a subset of {\bf R} and {\bf R} respectably. 
The symbol $\langle \ \rangle$ indicates the average over the posterior 
distribution.\\ 
An expression  $\arg \max_z f(z)$ indicates
a value of $z$ that maximizes $f(z)$ and \\ Kronceker delta $\delta_{w,z}$ 
is defined as usual, i.e., $\delta_{w,z}=1$ if $w=z$, else $\delta_{w,z}=0$.}

\begin{tabular}{@{}ccl}
\br
loss function $(L)$ & optimal estimator $(\hat{x})$ & comments \\ \mr
$\sum_i (x_i - \hat{x}_i)^2$ & $\hat{x}_i = \langle x_i \rangle$ &  \\ \mr
$\sum_i | x_i - \hat{x}_i |$ & $\hat{x}_i =$ median of $p_i(x_i)$$\, {}^{\rm a}$ &  
\\ \mr
$1 - \prod_i \delta_{x_i ,\hat{x}_i}$ & $\{\hat{x}_i\} = \arg \max_x p(x|y)$$\, 
{}^{\rm b}$ & $x_i$  : a discrete variable \\ \mr
$\sum_i (1-\delta_{x_i ,\hat{x}_i}$) & $\hat{x}_i = \arg \max_{x_i} p_i(x_i)$$\, 
{}^{\rm a}$ & $x_i$ : a discrete variable \\ \mr
$- \sum_i x_i \hat{x}_i$ & $\hat{x}_i =\frac{\langle x_i \rangle}{|\langle x_i 
\rangle|}$ & $\hat{x}_i \in {\pm 1}$ \\ \mr
$- \sum_i  \{x_i \log \hat{x}_i + (1-x_i) \log (1-\hat{x_i}) \}$ & $\hat{x}_i = 
\langle x_i \rangle$ & $0 < x_i, \hat{x}_i < 1 $ \\ 
\br
\\
\end{tabular} 
\\ 

 ${}^{\rm a}$ Here $p_i(x_i)$ indicates the marginal distribution $p_i(x_i)= 
\sum_{\{x_j\} \,  j \neq  i} p(x|y)$ of $x_i$, \\ \ \ where  $\sum_{\{x_j\} \, j \neq  
i}$ means the summation over $x$ with a fixed value of the $i$th component $x_i$. 
\\

${}^{\rm b}$ It is often called ``MAP (Maximum A Posteriori) estimator''. 

\endfulltable

So far, our discussion in this section does not depend on the notion of the gauge 
invariance. Correspondences between loss functions and optimal estimators shown in 
the Table 1 are independent of the existence of gauge invariance of the model.  With 
gauge invariance, we can prove an additional result.  Let us assume that the model 
is gauge invariant and the following properties of the loss function $L$ and the 
estimator $\hat{x}(y)$
\begin{equation}
\label{dCOV}
L(\, U_z (x), U_z (\hat{x}) \, ) = L( x, \hat{x} )
\end{equation}
\begin{equation}
\label{estCOV}
U_z ( \, \hat{x}(y) \, ) =\hat{x}(\, V_z(y) \,)
\end{equation}
are satisfied for all $z$ (The mappings $V_z$ and $U_z$ are defined in the section 
\ref{secN}.). 
Then, we can show that the expected loss $[ L(x,\hat{x}(y))]_{p(y|x)}$ with any fixed 
$x$ is independent of the value of $x$. The proof (\cite{Berger} p.396, \cite{Text4} 
p.168)
is as follows:
\begin{eqnarray}
\label{proof1}
[L( x, \hat{x}(y) )]_{p(y|x)} 
& & = \sum_y \, L(x,\hat{x}(y) ) \, \cdot \, p(y|x) \nonumber \\
& & = \sum_y \, L(x^*,U_z (\hat{x}(y))) \, \cdot 
\, p(V_z(y)|x^*) \nonumber \\
& & = \sum_y \, L(x^*,\hat{x}(V_z(y)))  \, \cdot 
\, p(V_z(y)|x^*) \nonumber \\
& & = \sum_y \, L(x^*,\hat{x}(y)) \, \cdot \, p(y|x^*) \nonumber \\
& & = [L( x^*, \hat{x}(y) )]_{p(y|x^*)} 
\end{eqnarray}
where $x^*$ is an arbitrary chosen ``standard'' configuration, say
a ferromagnetic state, and $z$ is chosen to satisfy 
the relation $x^* = U_z (x)$.
The result (\ref{proof1}) means that the estimator performs equally well for any value 
of the original parameter $x$. In terms of statistics \cite{Text4}, an estimator that 
is optimal within the class of the estimators 
with such uniformity is called a minimum risk 
equivariant estimator (MRE) ~\footnote{ 
The term ``invariant'' is also used. 
The author prefers ``covariant'', but does not know whether it has been used by 
statisticians. 
Here we restrict ourselves within
the special form of $U_z$ and $V_z$ induced by the gauge 
transformation group of Ising spin glass.
See the reference \cite{Berger,Text4} for definitions and results 
with an arbitrary group of transitive transformations.
}.
The case discussed in Ruj\'{a}n \cite{Rujan} and Nishimori 
\cite{Nishicode} corresponds to a special example of MRE.

In fact, we can remove the 
assumption (\ref{estCOV}) on the estimator, 
if the estimator is optimal and the optimal estimator is known to be unique. 
This means that if the loss function is gauge invariant,
the corresponding optimal estimator is 
automatically gauge covariant and 
satisfies (\ref{estCOV})~\footnote{ It is not true without an additional assumption
on the uniqueness. A counter example is given by a binary symmetric
channel with extreme noise $q=1/2$, which transmits no information. 
For this example, any estimator 
is ``optimal'' for $L(x,\hat{x})=-\sum_i \hat{x}_ix_i$. 
Some of them, say an estimator that returns a constant as estimates, are evidently 
not a MRE.}  . 
The proof is easy, if we note
that the estimator defined by
\begin{equation}
\hat{x}_z(y) = U_z^{-1} ( \, \hat{x}(V_z(y)) \, )
\end{equation}
is an estimator of the equal performance to the original
estimator $\hat{x}(y)$, i.e.,
\begin{equation}
\label{uniq}
[ [ L(x,\hat{x}_z(y))]_{p(y|x)}]_{\pi(x)} = [ [ L(x,\hat{x}(y))]_{p(y|x)}]_{\pi(x)} 
{}_.
\end{equation}
The relation (\ref{uniq}) is confirmed by
the calculation similar to that in the proof of (\ref{proof1})
under the assumption of (\ref{dCOV}) and the gauge covariance
of $p(y|x)$ and $\pi(x)$.
Thus, with the assumption of the uniqueness of the optimal estimator, 
$\hat{x}_z(y)$ should be coincides with $x(y)$
for any value of $z$. It proves the relation (\ref{estCOV}).

\label{A2}

\ack
The author would like to thank  Y.~Kabashima and D.~Saad for
fruitful discussions and kind advices.

\end{document}